# APP CREATION IN SCHOOLS FOR DIFFERENT CURRICULA SUBJECTS - LESSONS LEARNED

**Bernadette Spieler, Christian Schindler, Wolfgang Slany, Olena Mashkina**

*Graz University of Technology, Austria*

## Abstract

The next generation of jobs will be characterized by an increased demand for people with computational and problem solving skills. In Austria, computer science topics are underrepresented in school curricula hence teaching time for these topics is limited. From primary through secondary school, only a few opportunities exist for young students to explore programming. Furthermore, today's teachers are rarely trained in computer science, which impairs their potential to motivate students in these courses. Within the "No One Left Behind" (NOLB) project, teachers were supported to guide and assist their students in their learning processes by constructing ideas through game making. Thus, students created games that referred to different subject areas by using the programming tool Pocket Code, an app developed at Graz University of Technology (TU-Graz). This tool helps students to take control of their own education, becoming more engaged, interested, and empowered as a result. To ensure an optimal integration of the app in diverse subjects the different backgrounds (technical and non-technical) of teachers must be considered as well. First, teachers were supported to use Pocket Code in the different subjects in school within the feasibility study of the project. Observed challenges and difficulties using the app have been gathered. Second, we conducted interviews with teachers and students to underpin our onsite observations. As a result, it was possible to validate Pocket Codes' potential to be used in a diverse range of subjects. Third, we focused especially on those teachers who were not technically trained to provide them with a framework for Pocket Code units, e.g., with the help of structured lesson plans and predefined templates.

## 1 INTRODUCTION

Currently, we see three major problems in the Austrian computer science school curricula: 1) skills like computational thinking [1], programming, and game creation are not compulsory in high schools in Austria [2], 2) teachers do not have the time or the expertise to teach such skills [3], and 3) there is a lack of case studies and useful age-appropriate teaching material for teaching computer science. The computer science curriculum is only defined for the Year 9 [4]. This curriculum covers the basic competencies in dealing with technologies such as Office products or descriptions of learning about basic principles of automata, algorithms, and programs. Computer science classes for higher grades (Year 11 to Year 13) are offered as an optional subject [5]. In addition, teachers are insufficiently trained for applied computer science education because it is not seen as a major subject in their education [3]. The reason for this is that most universities do not have programs for computer science education to prepare the future generations of teachers, or there is little interest in these programs. In the year 2016, only 5 students enrolled in lectureship courses for computer science at the TU-Graz[1]. Moreover, for higher grades in Austrian schools, there is no officially authorized literature containing practical, state of the art examples for computer science education. Even so, there are currently efforts to reinvent the computer science schoolbook with a focus on practical examples and learning materials, licensed under Creative Commons – CC BY[2].

Teachers must generally stick to an often overburdened lesson plan and have little freedom in planning their lessons depending on the subject. In mathematics, for example, the few units available are a serious problem in Austria [3]. Thus, teachers often feel overwhelmed by the increasing amount of new media and different learning software products available. For this reason, an understandable and easy to use tool for teaching students computer science related skills is essential for both technical and non-technical teachers in Austria. Pocket Code fills this need by providing teachers with an interesting approach to introduce novice users to programming and conceptual thinking within the context of science, technology, engineering, arts, and mathematics (STEAM) related subjects. Therefore, Pocket Code is not limited to be used only in computer science, but also in physics, music, arts, or language learning courses.

---
[1] https://online.tugraz.at/tug_online/Studierendenstatistik.html
[2] https://learninglab.tugraz.at/informatischegrundbildung/index.php/oer-schulbuch/

This paper focuses solely on the perspective of the teachers and aims at providing guidance for those who start using Pocket Code in their lessons without any prior knowledge in programming. The paper is structured as follows: Section 2 shows the importance of teaching computational thinking skills in secondary schools through a literature review and points out how teachers can transfer these skills. Section 3 provides an overview of the EU NOLB project, the Pocket Code app, and holds a list of all courses held during the Austrian pilots. Section 4 summarizes the observed challenges gathered while working together with different types of teachers. Within Section 5, the results are discussed, and Section 6 concludes the paper with an outlook of work to be started in the future.

## 2   LITERATURE REVIEW: COMPUTATIONAL THINKING SKILLS

According to Wing [1] computational thinking means to:

- Understand different aspects of a given problem,
- Link computational tools and techniques to this problem,
- Understand the limitations and power of the used tool, and to
- Generalize and apply this solution process to a variety of problems

Learning how to program reinforces computational thinking skills [6]. However, it's important to consider that computational thinking is not just about programming. First, students should think about possible solutions to a given problem (use of problem solving skills) and then implement their ideas by using a computing device (use of programming skills) [7]. To successfully implement the solution students have to apply different programming concepts, such as loops and conditions, as well as practices, such as abstraction and debugging [8, 9].

There is already a small but consistent change in the educational system and in teachers' views regarding the set up of computer science courses [10, 11, 12, 13]. The focus on new initiatives lies on teaching fundamental principles and concepts, thinking abstractly, and working in multiple was of abstraction.

From the teachers' perspective, applying computational thinking skills to students is actually a challenging task [14]. The literature assumes [15, 16] that it is easier to combine concepts of algorithms and programming with activities which are engaging and fun but also relevant at the same time. Games are known as an effective approach for teachers to motivate students to interact and communicate as well as to learn [17]. Kafai [18, 19] points out that it is much more effective when students program games on their own instead of just learning about programming. This allows students to collaborate and construct solutions for problems. In addition, playful activities provide an authentic context in which subjects can be situated rather than just being taught. Through an interdisciplinary approach, teachers not only transfer their subject knowledge but also teach fundamental programming skills.

## 3   THE "NO ONE LEFT BEHIND" PROJECT: EXPERIMENTAL PILOTS

One of the goals of the ongoing European project "No One Left Behind"[3] (NOLB) is to integrate mobile programming [20] into different subject areas. The project team performed actions in primary through secondary levels in three phases (feasibility study, first, and second cycle) by setting up three European experimental pilots in Austria, Spain, and the UK. The NOLB project started in January 2015 and will end in June 2017. Programming was introduced via the learning environment Pocket Code. This section describes the Pocket Code app, and provides an overview about all courses and teacher characteristics of the Austrian pilots.

---
[3] http://no1leftbehind.eu/

### 3.1 Pocket Code: The mobile way of programming

Pocket Code[4] is a mobile app that allows users to create their own games, animations, interactive music videos, and many types of other apps directly on one's mobile device. Pocket Code uses the visual "lego"-style programming language Catrobat, which is very similar to Scratch[5]. The programming environment Scratch is already well known at schools and has been adopted into many computing classes all over the world [21]. The use of these visual programming languages keeps the focus on the semantics of programming and eliminates the need to deal with syntactical problems. Through programming with Pocket Code's visual programming language, teachers can gain insights on how students think because the tool visualizes cognitive processes automatically on the Pocket Code's community webpage[6]. On this page the code and code statistics for every submitted Pocket Code program can be inspected (see Figure 1).

```
Total number of SCRIPTS:    88
Total number of BRICKS:     553
Total number of OBJECTS:    37
Total number of LOOKS:      62
Total number of SOUNDS:     7
Total number of GLOBALS:    9
Total number of LOCALS:     34

EVENT BRICKS:     CONTROL BRICKS:   MOTION BRICKS:    SOUND BRICKS:
Total:     112    Total:     235    Total:     85     Total:     7
Different: 6      Different: 8      Different: 6      Different: 2

         LOOKS BRICKS:     PEN BRICKS:       DATA BRICKS:
         Total:     61     Total:     0      Total:     53
         Different: 6      Different: 0      Different: 3
```

*Figure 1. Code statistics on Pocket Code's sharing community platform provides teachers with an overview of the program*

Although, the schools participating in the pilot seemed to be sceptical at first about whether smartphones should be used in schools for teaching at all, they soon were convinced of the advantages of mobile devices for programming [22]. For instance, mobile phones are lighter, cheaper, easier to interact with, more portable, and easier for schools to maintain compared to PCs and laptops [23]. Hence, teachers need not to reserve computer labs for programming education. Furthermore, not all students have access to a PC for homework, but almost all of them own a smartphone.

### 3.2 The Austrian study: Selected curricula subjects

As part of the NOLB project, Pocket Code was introduced in Austrian partner schools in the following curricula subjects: physics, music, arts, language learning, and computer science. The Austrian pilot schools were situated in and around Graz. The preparation of the feasibility study took current curricula and lesson plans into account. For preparation, suitable classroom projects, use cases, and templates were created for use with Pocket Code in different courses. Eleven teachers participated on a voluntary basis in the Austrian study and were invited to use the material during their classes. They provided feedback for improvement in a later stage. These teachers, their subjects, background, and a description of how they used Pocket Code in their courses are visualised in Tables 1 to 6. All three participated schools are AHS – Academic Secondary Schools (for more information see [24]). Before the start of the feasibility study (autumn 2015), a first engagement workshop for students was organized. In addition, teacher meetings and training sessions were conducted to ensure a smooth start and to kick off the feasibility study. Moreover, an online questionnaire was set up to collect information about the teachers' digital skills and abilities (participating teachers: 1, 2, 3, 4, 7, 8, and 13, see tables below). This helped to gain better understanding of how resources such as training materials can be optimized to fit teachers' needs. The results of the questionnaire showed that all of them had already used digital media in their classes e.g., presentations, links, video clips, and all except two teachers (7, 8) also used boards/forums, or blogs and other tools for group discussion, or interactive elements such as mobile quizzes or polls. Five out of seven mentioned that they used the school's Management Information System (MIS) to electronically record and monitor information about student's attendance, behaviour, and achievements. Two teachers (1, 7) mentioned that they play computer games for fun and enjoyment and already used computer games in while teaching their classes.

---

[4] https://catrob.at/pc
[5] https://scratch.mit.edu/
[6] https://share.catrob.at/pocketcode/

During the feasibility study (FS), eleven courses were performed; during the first cycle (FC), two courses; and during the second cycle (SC), nine courses. Teachers could decide whether the students should work in small groups (SG) on a joint program (every student should then create one game level which at the end would be merged into one big game); do pair work (PW), where two students work on the same program (either at one or at two tablets); or work individually (I). In addition, teachers could ask the project team for assistance during their lessons for questions specific to Pocket Code. During the feasibility cycle, the team tried to take the role of observer rather than guide.

*Table 1. Teachers of pilot school one: AHS - A Bilingual school.*

| Teacher Nr. | Gender | Technical background | Subject in which Pocket Code has been integrated |
|---|---|---|---|
| 1 | female | no | English |
| 2 | female | no | English |
| 3 | male | yes | computer science |
| 4 | male | yes | computer science |
| 5 | female | no | Physics |
| 6 | male | no | fine arts |

*Table 2. Courses at pilot school one.*

| Course Nr. | Pilot phase | Teacher Nr. | Students | Topic and Learning Goal | Learning scenario |
|---|---|---|---|---|---|
| 1 | FS | 1, 2, 3, 4 | 48 Year 10 | Topic: book retelling<br>Goal: To program an interactive quiz game in computing to a book read in English class. | SG |
| 2 | FS | 4 | 6 Year 10 | Topic: Game design<br>Goal: Create a simple game (start, instruction, game, and end screen). | I |
| 3 | FS | 5 | 26 Year 8 | Topic: Density of objects and liquids<br>Goal: Create a game where objects glide according to their physical properties (density) and apply the formula. | PW |
| 4 | FS | 6 | 20 Year 10 & 20 Year 11 | Topic: Game design<br>Goal: Create a game and finish one level per student. | SG |
| 5 | SC | 6 | 75 Year 10 | Topic: Game design<br>Goal: Add a start and game over screen, avoid/ catch something, or tell a story and add an interactive part. | SG |
| 6 | SC | 5 | 26 Year 8 | Topic: Newton's 2nd law of motion<br>Goal: Create a game where objects glide according to their physical properties (mass, acceleration) and apply the formula.<br>Used template: Physical simulation | I |
| 7 | SC | 4 | 12 Year 10 | Topic: game design<br>Goal: Create an adventure RPG game and apply it to different subject areas e.g., biology, music, etc.<br>Used template: Adventure RPG | I |

*Table 3. Teachers of pilot school two: AHS.*

| Teacher Nr. | Gender | Technical background | Subject in which Pocket Code has been integrated |
|---|---|---|---|
| 7 | female | yes | physics, computer science |
| 8 | female | no | fine arts, computer science |

*Table 4. Courses of pilot school two.*

| Course Nr. | Pilot phase | Teacher Nr. | Students | Topic and Learning Goal | Learning scenario |
|---|---|---|---|---|---|
| 8 | FS | 8 | 29 Year 7 | Topic: Alice in Wonderland<br>Goal: Create a vocabulary game by adding missing parts within the code. | PW |
| 9 | FC | 7 | 12 Year 10 | Topic: Game design<br>Goal: Create a Quiz template for Year 7. | I |
| 10 | FC | 7 | 29 Year 7 | Topic: Structure of matter<br>Goal: add five questions to the quiz.<br>Used template: Quiz | PW / I |
| 11 | SC | 8 | 29 Year 8 | Topic: renaissance, baroque, and romanesque<br>Goal: Create a puzzle game with your graphics.<br>Used template: Puzzle | PW/I |
| 12 | SC | 8 | 12 Year 13 | Topic: game design<br>Goal: create an adventure game (start, game, end screen) and a quiz with 5 questions.<br>Used template: Quiz | SG / I |
| 13 | SC | 7 | 29 Year 8 | Topic: Physical experiments<br>Goal: Add an explanation and animation of a performed experiment.<br>Used template: interactive book | I |

*Table 5. Teachers of pilot school three: AHS with focus on computer science.*

| Teacher Nr. | Gender | Technical background | Subject in which Pocket Code has been integrated |
|---|---|---|---|
| 9 | male | yes | computer science |
| 10 | male | no | music |
| 11 | female | yes | computer science |

*Table 5. Courses of pilot school three.*

| Course Nr. | Pilot phase | Teacher Nr. | Students | Topic and Learning Goal | Learning scenario |
|---|---|---|---|---|---|
| 14 | FS | 9 | 13 | Topic: Quiz about computer science<br>Goal: Create a quiz. | PW |
| 15 | FS | 10 | 21 | Topic: Quiz about musical instruments<br>Goal: Add sounds and catch the used musical instruments. | PW |
| 16 | SC | 11 | 17 | Topic: Galaxy<br>Goal: Create an action game (start, game and end screen). | I / SG |

Teachers who had introduced this gamified approach during the feasibility study and the first cycle used the app Pocket Code. Teachers who had classes during the second cycle used the app Create@School. This special version was developed for use at schools. It integrates examples in the form of predefined templates, as well as advanced features that should simplify the programming experience for students. This is described in more detail in Section 4. The preparation, creation, and evaluation of these templates was part of the authors' previous work which is already submitted and currently under peer review.

## 4 RESULTS

This section summarizes feedback from teachers collected via interviews, on-site observations, and discussions. First, their experiences during the feasibility study and the first cycle have been summarized and compared with the students' outcomes. Based on these results, the following actions have been performed: the new school app Create@School has been developed and improvements for the setup of the courses have been implemented. Second, these efforts have been evaluated through usage by teachers and students in the second cycle.

The outcome of the feasibility study (students n=187) showed diverse results. Data was collected in two ways. First, the students took part in three quantitative surveys (before, during, and after using the app Pocket Code) to question their attitude towards the learning material and the application of Pocket Code in the lessons. Second, the learning objectives defined by the teacher beforehand were measured against the learning outcomes. By the end of the feasibility study, 105 out of 172 project submissions were rated clearly positive, meaning the students fulfilled the learning goal defined by the teachers. This shows the potential of Pocket Code to act as a supportive learning tool by leading to the accomplishment of academic curriculum objectives. To get feedback from teachers, they were asked at the beginning of the study to journalize each lesson. In addition, we had a debriefing with all teachers in March 2016 (before the start of the 1st cycle).

The evaluation of the student surveys and the teachers' feedback were not uniform but mixed throughout both groups (teachers and students), showing points of improvement for the use of Pocket Code at schools. Suggestions for improvement not only include the application itself, but also the style of teacher training and support, preparation of tutorials and lesson content and the backing of the courses. Consequently, only seven teachers (teacher 4, 5, 6, 7, 8, 11 and 13) decided to continue with the project during the cycles (FC, SC). Teacher 1 and 2 (who were both working on the same projects, both non-technical) were disappointed about the students' outcomes and expected more advanced games from their students. As a result, the students felt quite stressed during the course, thus leading to a bad experience for them as well. These two teachers never programmed with Pocket Code on their own, showed no interest in learning about programming, and expected the students to find out everything on their own, e.g., search for help online. In addition, the 15 to 16-year-old students are used to relatively sophisticated games (e.g., World of Warcraft[7]). With Pocket Code, however, they have to downgrade their expectations and hence their implementation that the games become curiosities. Teacher 3 and 9 both have a technical background and are advanced programmers. Teacher 3 had already programmed with Scratch, but at that time Pocket Code was not fully Scratch compatible and lacked comfort-functionality like automatic collision detection. Both initially had high expectations towards the app but were eventually disappointed as well.

The remaining group of teachers was highly motivated to work with the app during the experimental cycles. These teachers had either more than one class at one time (e.g., teacher 6), several classes in different subjects (e.g., teacher 7 and 8), or the same class in several of the same subjects (teacher 6, 7 and 8). However, they also said that certain parameters need to be established in order to guarantee future success of the units. Teacher 6 evaluated the Pocket Code exercise of his students precisely [25]. His recommendations included to

1) Design a well-structured pedagogical framework to avoid an overload of technical complexity,

2) Prepare a hardcopy handbook that promotes the overview of all bricks and their functions, along with frequently used brick combinations, and to

3) Limit the project duration to 4-5 double lessons in order to concentrate work towards a deadline and prevent student fatigue with the project.

Based on these results, the focus during the first project cycle was twofold. First, the team decided that the app should become Scratch compatible with a main focus on usability and feature completeness. As a result, 47 new features were developed during the first cycle, which lead to a specialized app for use in schools with the name Create@School. This special flavor of Pocket Code was published in September 2016 as a beta version at Google Play for use at our pilot schools during the second cycle. Second, the team developed more appropriate teacher guidelines, predefined game templates, and resources for teaching. In particular the team provided 1) very general tutorials, e.g., beginner steps, brick documentations, and video tutorials[8], 2) specific templates for certain lessons/genres (e.g., action, adventure, puzzle, and quiz) available within the menu of Create@School, and tutorials for teachers, and 3) more mandatory trainings for all teachers, as well as one-to-one trainings to prepare specific units. This should allow teachers to feel better prepared for their lessons by reassuring them that they will be able to conduct their lessons without assistance, regardless their technical background.

---

[7] https://worldofwarcraft.com

During the second cycle (n=160), teachers used predefined templates for their courses (teacher 3, 5, 7 and 8). Teacher 6 again used a package of laminated analogue cards of the Pocket Code bricks, first to translate their composed narratives into coding threads; in a second step, students programmed a game from scratch. Students in the course of teacher 11 started with their own action games. For a preliminary evaluation, the data of six courses (4 teachers) has been observed. Two teachers used a template in their courses (n=47, Year 8), and two teachers told their students to work in small groups and to start a game on their own (n=79, Year 10). This time, 31 students who used a template achieved the learning goal (16 students did not) and only 38 students who started with their own programs achieved the goal (41 students did not). This ratio of successfully reaching the learning goal to failing, clearly shows the advantage of using templates instead of developing from scratch. The feedback and experience of teacher 6, who did not use a template (but used prepared material for guidance, e.g., tutorial cards), was very similar to his previous courses during the feasibility study and the students again needed a lot of guidance. On-site observations led to the conclusion that the students did not properly define the goals of their games and hence lost focus. Additionally, they had only two double units instead of four, like in the feasibility study, for finalizing their games. Consequently, these students were once again frustrated, stressed, and not able to finish their games (thus they did not reach the learning goal). Teacher 6 felt discouraged as well.

The rest of the teachers felt more confident during the lessons due the use of predefined templates or the tutorial cards. On-site observations showed that the students had less questions, were very concentrated, worked on their own, solved the underlying problems, and felt engaged with the whole class. Since everybody had the same learning goal, they also had similar problems and in that way could help each other.

All seven teachers (and additionally teacher 11) are planning to continue working with Pocket Code after the NOLB project and plan to integrate Pocket Code permanently in their lessons.

## 5  DISCUSSION

The preliminary evaluation of the created games showed that many important concepts necessary for game-development are now easier to understand for the target group, e.g., accurate collision detection between objects, or speech bubbles for storytelling games. In addition, the predefined templates were adapted to different subjects. The collected results showed Pocket Code's potential to be used in diverse subjects, e.g., a quiz template used in physics as well as in arts. Additional resources and better training allowed students to focus more on the subject-relevant problem solving activities than on understanding the whole functionality of the app. This directly led to more time to express their creativity on different levels and more time for extra tasks.

Teachers who used the templates with a given learning goal needed to adopt them to their subject. On one hand, they needed more time for preparing their courses or they needed more individual meetings with the NOLB team. On the other hand, the classroom atmosphere was much more relaxed because teachers guided the students more by focussing on the topic rather than explaining complex program structures to them. Most teachers (5, 7, 8 and 13) switched from group or pair work to individual work. The reasoning was that individual work fosters computational thinking. If everybody is working on the same problem, everybody could find a different solution for it. In this way, students are able to support each other, working independently but in small groups, thus feeling more engaged as a result. Furthermore, it has been observed that if everyone is working on one level of a game (and on their own problem), much more stress arises because everyone has to work on his or her own problem in order to merge it together into one game at the end.

---

[8] https://edu.catrob.at/

The observations showed three similar challenges for teachers:

- their confidence in teaching computing as a subject,
- structure of the course and defining the learning goal, and
- the issue of having enough technical support in the classroom.

In order to give teachers more confidence and guidance in future, it has been planned to create a hardcopy book as proposed by teacher 6. These strategies could help teachers to feel more confident in using Pocket Code in the classroom.

## 6  CONCLUSION AND FUTURE WORK

This paper summarizes teachers' experiences with the educational app Pocket Code, shows examples of how to integrate the tool in different subjects, and delivers strategies on how to optimize support for teachers with different backgrounds in order to use the app. To evaluate the effectiveness of these efforts, a survey will be conducted by using a tool which provides usability and design evaluation [7] to see if Create@School is now better suited for schools. For the future, follow up questions regarding the material and guidelines will be collected as well. Results will be available after the last programming units at the end of May 2017.

To help teachers in the upcoming assessment and feedback process of students' projects, a Project Management Dashboard (PMD) and tracking of analytics data, including both qualitative and quantitative data have been developed. First, all events during programming were tracked within the Create@School app. Actions like create a new program, add a brick, time spent with programming, or the use of game templates are tracked and stored in a database. The app uses the commercial BDSClientSDK (Big Data Services Client software development kit), which is a very simple to use and lightweight library without external dependencies that allows developers to send different types of events related with their applications to the Big Data Services (BDS) platform. In that way, it is possible to explore information about users, their sessions, and their actions. Second, teachers are able to assess the pupil's programs within the PMD. The PMD provides the framework needed to manage students and classes for each teacher of the NOLB project. Additionally, the PMD allows qualitative evaluation of students regarding the completion of projects and achievements of objectives. Both services were integrated during the project's first and second cycle and applied in schools from April 2017 until the end of the project in June 2017.

## ACKNOWLEDGEMENTS

This work has been partially funded by the EC H2020 Innovation Action No One Left Behind, Grant Agreement No. 645215.

## REFERENCES

[1]   J.M. Wing, "Computational thinking". In Commun. ACM, vol. 49, no. 3, pp. 33-35, 2006.

[2]  Federal Ministry of Education Austria 1 (2017) Highscool Curricula (BMB - Bundesministerium fuer Bildung: Lehrpläne der AHS Oberstufe), [online], https://www.bmb.gv.at/schulen/unterricht/lp/lp_ahs_oberstufe.html.

[3]  T. Jarz, "Current problems in school informatics". University of Applied Sciences Styria ("Aktuelle Probleme in der Schulinformatik". Pädagogische Hochschule Steiermark.), pp. 116-120, 2016. Accessed 10 May, 2017. Retrieved from: https://www.ahs-informatik.com/tagungsband-25-jahre-schulinformatik/informatik-ein-fach/

[4]  Federal Ministry of Education Austria 2, Highschool Curricula - Informatics mandatory 9th grade, (BMB - Bundesministerium für Bildung: Lehrpläne der AHS Oberstufe - Informatik 5. Klasse), Accessed 10 May, 2017. Retrieved from: https://www.bmb.gv.at/schulen/unterricht/lp/lp_neu_ahs_14_11866.pdf.Adrian

[5]  Federal Ministry of Education Austria 3, highschool Curricula - Informatics optional 10th-12th grade, (BMB - Bundesministerium für Bildung: Lehrpläne der AHS Oberstufe - Informatik 6.-8. Klasse), Accessed 10 May, 2017. Retrieved from: https://www.bmb.gv.at/schulen/unterricht/lp/lp_neu_ahs_14_11866.pdf.

[6]   A. Domínguez, „Gamifying learning experiences: Practical implications and outcomes" , In Computers & Education, vol. 63, pp.380–392, 2013.

[7]   C.C. Selby, "Promoting computational thinking with programming", In Proceedings of the 7th Workshop in Primary and Secondary Computing Education (WiPSCE '12), pp. 74-77, 2012.

[8]   S.Y. Lye and J.H.L. Koh, "Review on teaching and learning of computational thinking through programming: What is next for K-12?", in Computers in Human Behavior, vol. 41, pp. 51-61, 2014.

[9]   Y. Kafai and Q. Burke, "Computer programming goes back to school", in Phi Delta Kappan, vol. 95, no. 1, pp. 61, 2013.

[10]  A. Repenning, D.C. Webb, K.H. Koh, H. Nickerson, S.B. Miller, C. Brand, I. Horses, A. Basawapatna, F. Gluck, R. Grover, K. Gutierrez, and N. Repenning, "Scalable Game Design: A Strategy to Bring Systemic Computer Science Education to Schools through Game Design and Simulation Creation", in Trans. Comput. Educ., vol. 15, no. 2, Article 11, pp. 1-31, 2015.

[11]  K.H. Koh, A. Basawapatna, V. Bennett, and A. Repenning, "Towards the Automatic Recognition of Computational Thinking for Adaptive Visual Language Learning", In Proceedings of the 2010 IEEE Symposium on Visual Languages and Human-Centric Computing (VLHCC '10), IEEE Computer Society, pp. 59-66, 2010.

[12]  J. Walden, M. Doyle, R. Garns, and Z. Hart, "An informatics perspective on computational thinking", In Proceedings of the 18th ACM conference on Innovation and technology in computer science education (ITiCSE '13), ACM, pp. 4-9, 2013.

[13]  A. Balanskat and K. Engelhardt, "Computing our future: Computer programming and coding- Priorities, school curricula and initiatives across Europe", European Schoolnet (EUN Partnership AISBL), 2014.

[14]  S. Sentence and A. Csizmadi, "Teachers' perspectives on successful strategies for teaching Computing in school", IFIP TC3 Working Conference "A New Culture of Learning: Computing and Next Generations,  pp. 1-10, 2015.

[15]  T.J. Kopcha, L. Ding, K.L. Neumann, "Teaching Technology Integration to K-12 Educators: A 'Gamified'Approach", in TechTrends , vol. 60, no. 1, pp 62–69, 2016,

[16]  Z.B. Syamsul and S. Norshuhada, "Adapting Learning Theories in Mobile Game-Based Learning", In Digital Game and Intelligent Toy Enhanced Learning (DIGITEL), 2010 Third IEEE International Conference, pp. 124-128., 2010.

[17]  Y. Kafai and V. Vasudevan, "Hi-Lo tech games: crafting, coding and collaboration of augmented board games by high school youth", In Proceedings of the 14th International Conference on Interaction Design and Children (IDC '15), ACM, pp. 130–139, 2015.

[18]  Y. Kafai, "Playing and making games for learning: Instructionist and constructionist perspectives for game studies", In Games & Culture, vol. 1, no. 1, pp. 36-40, 2006.

[19]  S. Papert. "Mindstorms. Children, Computer, and Powerful Ideas", Basic Books, Inc., 1985.

[20]  S.B. Zaibon and N. Shiratuddin, "Towards Developing Mobile Game-Based Learning Engineering Model", in Computer Science and Information Engineering, vol. 7, pp. 649-653, 2009.

[21]  O. Meerbaum-Salant, M. Armoni, and M. Ben-Ari, "Learning computer science concepts with scratch", In Proceedings of the Sixth international workshop on Computing education research (ICER '10), ACM, pp. 69-76, 2010.

[22]   A. Botha, M. Herselman, and D. van Greunen, "Mobile user experience in a mlearning environment", In Proceedings of the 2010 Annual Research Conference of the South African Institute of Computer Scientists and Information Technologists (SAICSIT '10), ACM, pp. 29–38, 2010.

[23]  Forecast of the smartphone user penetration rate in Austria 2014-2021. Accessed 10 April, 2017, Retrieved from: https://www.statista.com/statistics/567976/predicted-smartphone-user-penetration-rate-in-austria/

[24] Federal Ministry of Education Austria 4, Education in Austria 2016/2017, Accessed 10 May, 2017. Retrieved from:
https://www.bmb.gv.at/enfr/school/bw_en/bildungswege2016_eng.pdf?5te5kh

[25] S. Wardell, "An Investigation of Creative Environments in the Western World", in Lambent Interfaces, 2016.

BibTex entry:

```
@InProceedings{SPIELER2017APP,
author = {Spieler, B. and Schindler, C. and Slany, W. and Mashkina, O.},
title = {APP CREATION IN SCHOOLS FOR DIFFERENT CURRICULA SUBJECTS - LESSONS LEARNED},
series = {9th International Conference on Education and New Learning Technologies},
booktitle = {EDULEARN17 Proceedings},
isbn = {978-84-697-3777-4},
issn = {2340-1117},
doi = {10.21125/edulearn.2017.2315},
url = {http://dx.doi.org/10.21125/edulearn.2017.2315},
publisher = {IATED},
location = {Barcelona, Spain},
month = {3-5 July, 2017},
year = {2017},
pages = {5814-5823}}
```